\documentstyle[preprint,epsfig,aps]{revtex} 

\draft

\begin{document}

\title{Effective action for the homogeneous radion in brane cosmology}
\author{David Langlois, Lorenzo Sorbo}
\address{Institut d'Astrophysique de Paris, \\
(Centre National de la Recherche Scientifique)\\
98bis Boulevard Arago, 75014 Paris, France}
\date{\today}

\maketitle

\def\beq{\begin{equation}}

\def\eeq{\end{equation}}

\def\y{{\tilde y}}
\def\A{{\cal A}}
\def\R{{\cal R}}
\def\V{{\cal V}}
\def\B{{\cal B}}
\def\C{{\cal C}}
\begin{abstract}

We consider cosmological two-brane models with AdS bulk,  for which the radion,
i.e. the separation between the two branes, is time dependent.  In the case of 
two de Sitter branes (including Minkowski branes as a limiting case),  we
compute explicitly, without any approximation, the effective  four-dimensional
action  for the radion. With the scale factor on-shell,  this provides  the
non-perturbative dynamics for the radion. We discuss  the differences  between
the dynamics derived from the four-dimensional action with the scale factor
off-shell and the true five-dimensional dynamics.

\end{abstract}

\section{Introduction}

Intense activity has followed the recent suggestion that our universe could be
embedded in a higher dimensional spacetime, while exhibiting the usual  law of
gravity at least in the range of  scales which have been probed by gravity 
experiments. In the particular case of a single extra dimension, a lot of
research has been inspired  by the  Randall-Sundrum models \cite{rs99a,rs99b}
based on branes in an Anti de Sitter (AdS) five-dimensional  bulk spacetime. 
It has been shown explicitly in \cite{gt99} that usual gravity  (up to
corrections of  order $\mu^2 r^2$, where $\mu$ is the  AdS mass scale) is
indeed recovered in the  single brane model whereas,  in the two-brane models,
one gets  a Brans-Dicke type gravity.  This cannot be  compatible with
observations if we live  on the negative  tension brane (the case of interest
to solve the  hierarchy problem)  unless one invokes  a stabilization mechanism
for the radion,  i.e. the interbrane separation, such as the one suggested in 
\cite{gw99}. Whatever the specific mechanism,  the outcome is usually presented
as producing an effective potential  for the radion   which can be seen  as a
four-dimensional scalar field. 

In the cosmological context, it has been possible to solve exactly the
five-dimensional  Einstein equations when the bulk includes energy only in the
form of  a cosmological constant \cite{bdel99}.   Whatever the relative motion
of the branes, the radion does  not appear explicitly since the expansion law
of each brane is given,  independently of the other, by the unconventional
Friedmann equation (which follows from the Israel junction conditions) 
\beq
H^2={\kappa^4\over 36}\,\rho^2-\mu^2+{\C\over a^4},
\label{fried}
\eeq
where $H$, $a$ and $\rho$ are, respectively, the Hubble parameter, the scale
factor and the total energy density of each brane, while $\C$~is an integration
constant (analogous to the Schwarzschild mass). 

At first sight, the two results mentioned above do not seem to be compatible:
on the one hand, a four--dimensional perspective  yielding  a Brans-Dicke type
gravity with the radion in the r\^ole of the Brans-Dicke  scalar field, on the
other hand a  five--dimensional  analysis showing that the  Friedmann equation
in our brane is independent  of the radion. 

The purpose of the present work is to present a detailed analysis of the 
dimensional reduction of a two-brane model at the level of the variational 
problem, i.e. starting from the five-dimensional action and integrating  over
the extra-dimension to obtain an effective four-dimensional action. We restrict
our analysis to homogeneous systems, but we do not require any  restriction on
the radion velocity (spacetime fluctuations of the radion,  although at the
linearized level, have been considered in \cite{cgr99}  for Minkowski and
in~\cite{radion} for  de Sitter branes).

Equipped with the full effective four-dimensional action for homogeneous
fields, our analysis is then twofold. We first focus on the dynamics of the
radion in the cosmological background  given by the unconventional Friedmann
equation.  Since  we have not assumed the radion velocity to be small, we get
an action which includes the full nonlinear dynamics of the radion, and from
which we can  recover the equations of motion for the radion obtained in
\cite{bdl01}  by writing the junction conditions for a moving brane. 

We then consider the four-dimensional action as a variational problem  for the
full system (radion plus gravity and not only the radion),  and explore in
which regimes  the resulting dynamics is a good approximation of the true
five-dimensional  dynamics. Since  the latter is exactly known,  we can 
quantify  the deviation between the true and the  ``effective'' dynamics. Our
results illustrate the dangers of extending the four-dimensional intuition to
systems which are  intrinsically five-dimensional, as already pointed out  in
\cite{kkl01}. In some sense, our analysis enables us to go beyond the moduli
approximation (which consists in promoting free parameters of degenerate
solutions into  four-dimensional fields) used recently in \cite{kost01} in the
context of brane cosmology,  and provides a quantitative delimitation of  its
range of validity. 

Our plan is the following. We start, in section 2, with a description of the 
model and the definition of our coordinate system. In section 3, we  compute
the effective action by integrating explicitly over the extra  dimension. The
following section analyses the resulting dynamics of the  radion. Section 5 is
devoted to a comparison between the four-dimensional  effective dynamics and
the true dynamics. And we  give our conclusions  in the final section.

\section{Bulk metric}

We consider a portion of the five-dimensional Anti-de Sitter spacetime with
cosmological constant $\Lambda\equiv-6\,\mu^2$,  bounded by  two ``parallel'',
spatially homogeneous and isotropic three-branes. The fifth dimension is made
effectively periodic by   assuming a mirror (orbifold) symmetry across each of
the branes. 

The purpose of this paper is to derive the four-dimensional effective theory
for this system from the point of view of an observer in the brane
corresponding  to our universe, which we call $\B_0$. For this reason,  rather
than using  a coordinate system in which the metric is manifestly
static~\cite{kraus}, we will  prefer to use  a Gaussian normal (GN) coordinate
system based on $\B_0$, in which the metric has the form 
\beq
ds^2=g_{AB} dx^A dx^B=
-n(t,\y)^2\, dt^2+a(t,\y)^2\, \delta_{ij}dx^i dx^j+ d\y^2, \label{GNmetric}
\eeq
and where our brane-universe $\B_0$ is always at $\y_0=0$. We will moreover
assume that the   energy densities in $\B_0$ and in the second brane $\B_1$,
respectively  $\sigma_0$  and $\sigma_1$, are constants. We thus avoid the
delicate point of defining an action for a brane  with a generic perfect fluid 
as matter. 

The general (cosmological) solution to the Einstein equations for the above 
system, in the  GN coordinates,  is well known~\cite{bdel99}. In what follows,
we will assume that the  Schwarzschild-type constant ($\C$ in
eq.~(\ref{fried})) is zero. This means that we choose the bulk to be strictly 
AdS rather than Schwarzschild-AdS~\cite{kraus}. This choice simplifies the 
expression for the bulk metric, that acquires the form
\beq
n(t,\y)=N(t)\,\A(\mu \y)\,, \quad a(t,\y)=a_0(t)\,\A(\mu\y)\,, 
\quad \A(\xi)\equiv \cosh\xi -\eta_0 \,\sinh |\xi|\,\,,
\label{A}
\eeq
where we have introduced the dimensionless quantity
$\eta_0=\kappa^2\sigma_0/(6\mu)$ related to the energy density in our 
brane-universe. In an analogous way, we define 
$\eta_1=\kappa^2\sigma_1/(6\mu)$.

We will allow here  the second brane to  move with respect to the frame defined
by eq.~(\ref{GNmetric}). Its  position at any time $t$ will be given by  
\beq
\y_1=\R(t), 
\eeq  
where the function $\R(t)$ represents the  (homogeneous) radion, which  is thus
defined as the proper distance  between the two branes  in the GN coordinate
system  defined by eq.~(\ref{GNmetric}). We will always assume that  the GN
coordinate system does not break down before reaching the second brane, which
means that $\B_1$ is within the horizon.

Usually one prefers to express the size of the extra dimension,  and thus its
time dependence, in the metric components rather  than in a time-dependent
coordinate for the boundary of the extra-dimension. We can proceed similarly by
introducing a new  coordinate $y$ defined  as
\beq
\y=y\, \R(t),
\eeq
such that  the two branes are now at fixed positions, respectively $y_0=0$ and 
$y_1=1$. The metric (\ref{GNmetric}) now reads 
\beq
ds^2= -\left(n^2 - \dot\R^2 \, y^2\right)\,dt^2+a^2\,d{\bf x}^2 
+ 2\,\R\, \dot\R\,  y\, dy\, dt+\R^2\, dy^2\,\,. \label{metric}
\eeq
One can notice that, not only the metric component along the fifth dimension is
explicitly time dependent, but off-diagonal components also appear. This
explicit dependence of the metric on the radion velocity is usually claimed to
be ignorable  under the assumption that this velocity is  small. As we will see
later, this is justified only in very specific regimes.

\section{Effective action}

In this section we derive the effective four-dimensional action of the above
system as seen by an observer on $\B_0$.  The total five-dimensional  action 
includes  the action for each  brane, the bulk Einstein-Hilbert action (with a
cosmological constant term), but also  an extra  term, usually called the
Gibbons-Hawking \cite{gh} term  (involving the trace $K$ of the extrinsic
curvature tensor on the space boundaries), in order to take proper care  of
the boundary terms at $y=0$ and $y=1$. The total five-dimensional action thus
reads
\beq
S={1\over 2\kappa^2}\int d^5x\sqrt{-g}\,
\left({}^{(5)}R-2\,\Lambda\right)
+{1\over\kappa^{2}}\sum_{a=0,1}\int d^4 x\,\sqrt{-h_a} K
-\sum_{a=0,1}\sigma_a \int d^4 x\,\sqrt{-h_a},
\label{5D_action}
\eeq
where $h_a$ denotes the determinant of the induced metric on each of the 
branes. Substituting the metric ansatz (\ref{metric}) in this five-dimensional
action we get a functional of $n(t,y)$, $a(t,y)$ and $\R(t)$.

The four--dimensional effective action is defined as the result of the
integration of the above expression (\ref{5D_action})   over the fifth
dimension. This requires knowledge of the explicit dependence on $y$ of each 
term in the five-dimensional  action.  It is thus necessary  at this stage to
replace the metric components  $n(t,y)$ and  $a(t,y)$ by their  explicit form
given in  (\ref{A}), leaving  the  variables $a_0(t)$, $N(t)$ and $\R(t)$ as
unspecified functions of time.  The resulting four dimensional effective action
can be expressed  as
\begin{eqnarray}
S=\frac{1}{\kappa^2\,\mu}\int d^4x\,Na_0^3\,&& \left[{}^{(4)}R\,\psi_1(\R)
+12\,\mu^2\,\left(\psi_2(\R)+\psi_3(\R)\right)+\right. \cr
&&\left.+3\,\mu\,\A_1^3\,\left(\frac{\dot a_0}{N\, a_0}+\mu\,{\A_1'\over \A_1}\frac{\dot\R}{N}\right)\,
\ln\left({N\,\A_1-\dot\R\over N\,\A_1+\dot\R}\right)+6\,\mu\,\A_1^2\,\frac{\dot a_0}{N\, a_0}\,\frac{\dot{\R}}{N}+\right.\cr 
&&\left.-\kappa^2\,\mu\,\sigma_0
- \kappa^2\,\mu\,\sigma_1\,\A_1^4\sqrt{1-\frac{\dot\R^2}{N^2\,\A_1^2}}\, 
\right]\,\,,
\label{action}
\end{eqnarray}
with $\A_1\equiv \A(\mu\R)$~and $\A'_1\equiv \A'(\mu\R)$, while the $\psi$'s
are dimensionless  functions of $\mu\R$ defined by
\beq
\psi_1(\R)=\int_0^{\mu\R} d\xi\, \A\left(\xi\right)^2, \,\,\,
\psi_2(\R)=\int_0^{\mu\R} d\xi\, \A\left(\xi\right)^2\,\A'\left(\xi\right)^2, 
\,\,\,
\psi_3(\R)=\int_0^{\mu\R} d\xi\, \A\left(\xi\right)^4\,\,,
\label{psi}
\eeq
and ${}^{(4)}R$ is the (homogeneous) four--dimensional Ricci scalar
\beq
{}^{(4)}R=\frac{6}{N^2}\,\left(\frac{\dot{a}_0^2}{a_0^2}+\frac{\ddot{a}_0}{a_0}-
\frac{\dot{a}_0}{a_0}\,\frac{\dot{N}}{N}\right)\,\,.
\eeq
It can be useful to integrate (\ref{action}) by parts in time in order  to get
rid of  $\ddot{a}_0$ and $\dot N$.

The action (\ref{action}) is the main result of this work.  It depends on  the
three time-dependent functions  $N(t)$, $a_0(t)$ and $\R(t)$ and it contains, 
{\sl without any approximation}, the full  dynamics  of the radion, as will be
checked explicitly in the next section.  This result shows that if the radion
can indeed be considered  as a scalar field from the four-dimensional point of
view, its full nonlinear dynamics is governed by a rather unfamiliar type of
action.

\section{Equations of motion}

As the analysis carried out in the next section will show in detail, the
four--dimensional effective action~(\ref{action}), if  considered as embodying
the dynamics of the variables  $N(t)$, $a_0(t)$ and $\R(t)$, will not yield, in
general, the exact dynamics both of the radion and of the scale factor on our
brane. Indeed,  the dynamics of the latter is  governed by the five-dimensional
Einstein equations, and their content is  partially lost in the
four-dimensional reduction of the action.  In particular, the unconventional
Friedmann equation (equation (\ref{fried}) with $\C=0$), 
\beq
H_0^2\equiv\left(\frac{\dot{a}_0}{N\,a_0}\right)^2={\kappa^4\over 36}\,
\sigma_0^2-{\mu^2}= \mu^2\,\left(\eta_0^2-1\right),
\label{fried0}
\eeq
which characterizes  the evolution of the scale  factor in our brane-universe
$\B_0$, does not follow   from the four-dimensional action (\ref{action}).

However,  once we put gravity  {\it on-shell}, i.e. once we assume that the
scale factor is a solution  of (\ref{fried0}), the effective four-dimensional
dynamical equations  will be able to  yield  the exact dynamics of the radion
field.  We will now  show that this is indeed the case by  comparing the
equations of motion obtained from the variation  of the action with respect to
$N$ and $\R$ with the equations of motion   for the radion  obtained in
\cite{bdl01} directly from the junction conditions. 

As usual, the lapse function  $N$ is not a physical degree of freedom as it
corresponds to the arbitrariness in the definition of time. The variation   of
the action with respect to it yields a first integral, which corresponds in
ordinary cosmology to the (first) Friedmann equation. In our case the Friedmann
equation reads
\beq
H_0^2\,\psi_1+2\mu^2\left(\psi_2+\psi_3\right)+
\mu\left(H_0+\mu\,{\A_1'\over \A_1}\,\dot \R\right){\A_1^2\,\dot\R\over 1-(\dot\R/\A_1)^2}
={\kappa^2\mu\over 6}\,\sigma_0+{\kappa^2\mu\over 6}\,\sigma_1{\A_1^4\over 
\sqrt{1-(\dot\R/\A_1)^2}}, \label{friedmann}
\eeq
where we have set $N=1$ after variation. 

If we now impose that the expansion rate of $\B_0$ is given by
eq.~(\ref{fried0}), then the constraint (\ref{friedmann}) simplifies to give
\beq
\mu\,{\A_1'\over \A_1}+H_0\,{\dot\R\over\A_1^2}=\mu\,\eta_1\,
\sqrt{1-{\dot\R^2\over\A_1^2}}\,\,,
\label{bdl}
\eeq
where we have used the identities~$\left( \eta_0^2 - 1 \right) \, \psi_1 =
\psi_2 - \psi_3\,$,~$3 \, \psi_2 + \psi_3 = \A_1^3 \,\A_1'
+\eta_0\,$,~and~$\A'{}^2-\A^2=\eta_0^2-1$, which follow from the
definitions~(\ref{A})~and~(\ref{psi}). This equation  corresponds exactly to
the result  of  \cite{bdl01}, obtained   by writing directly the junction
conditions for a moving brane. 

The equation of motion for the radion itself, obtained by variation of the 
action (\ref{action}) with respect to $\R$, looks at first rather cumbersome, 
but can be remarkably simplified, using (\ref{bdl}), to yield finally
\beq
\ddot \R+3\,H_0\left(1-{\dot\R^2\over\A_1^2}\right)\dot\R
+\mu\,\A_1\,\A'_1\,\left(4-5{\dot \R^2\over \A_1^2}\right)=4\,\mu\,\A_1^2\,\eta_1
\left(1-{\dot\R^2\over\A_1^2}\right)^{3/2},
\eeq
which is also in agreement with the results of \cite{bdl01}. We can therefore
conclude that, if we impose the condition~(\ref{fried0}),  we recover the exact
dynamics of the radion.  As we will discuss in the next section, however, the
condition~(\ref{fried0})  does not emerge from the dynamical equations of the
four dimensional system.

As a particular situation, one can consider the case where the second brane is
not moving, i.e.  $\dot\R=0$. This implies, in the case of two de Sitter 
branes, that the second brane must be located at the equilibrium
position~$\R_{eq}$, defined by the condition 
\beq
\left({\A_1'\over \A_1}\right)_{\R=\R_{eq}}=\eta_1\,\,.
\label{equilibrium}
\eeq
The linearized dynamics around this equilibrium position  can be obtained from
the action  eq.~(\ref{action})  by substituting 
$\R=\R_{eq}+\delta\R\left(t\right)$,  and keeping terms up to the second order
in the perturbation  $\delta\R$. It is then easy to obtain   the canonically
normalized radion  $\varphi_c$ as
\beq
\varphi_c\simeq \sqrt{-\sigma_1}\,\A_1\,\delta\R\,\,.
\eeq
The effective mass can also be read from the second-order action and  one
recovers the familiar result $m^2_{eff}=-4\,H_0^2$.

As we mentioned  at the end of section 2,  ignoring  the dependence of the
metric (\ref{metric}) on $\dot\R$, as is usually done in the moduli
approximation, leads after expansion in powers of $\dot{\R}$  to a different
action for the radion, difference which shows up in  the  $\dot\R^2$ term. 
One can show that the discrepancy is negligible, in the regime near 
equilibrium,  if the condition  
\beq
\eta_0^2-1 \ll \frac{\A_1^2}{\mu\,\R}\,\frac{\kappa^2\,\vert\sigma_1\vert}{\mu}
\,\,
\eeq
is satisfied. This turns out ot be the case for the regime $\eta_0=1$
considered in~\cite{kost01}.

We finally notice that it is possible to  rewrite eq.~(\ref{bdl}) as an
equation for the scale factor  on the brane $\B_1$, which reads
\beq\label{fried1}
H_1^2=\mu^2\left(\eta_1^2-1\right),
\eeq
where the Hubble parameter on the second brane is given in terms of  $a_0$ and
$\R$ by the expression
\beq
H_1={1\over
\A_1\,\sqrt{1-{\dot\R^2\over \A_1^2}}}\,\left(H_0+\mu\,{\A_1'\over \A_1}\right).
\eeq
Two equivalent descriptions of the two-brane system are thus possible. One
consists in parametrizing the two branes by their scale factor  $a_0$ and $a_1$
respectively, in which case the dynamics is described by the two
unconventional  Friedmann equations~(\ref{fried0}) and~(\ref{fried1}), which
are completely independent.  The second description, directly related to the 
four-dimensional effective point of view, consists in choosing the scale factor
$a_0$ and the radion as degrees of freedom of the theory. The five-dimensional
setup can then be ignored, the radion appearing as a four-dimensional scalar
field in our usual four-dimensional spacetime,  but the memory of the
five-dimensional setup, which is embodied in the  unconventional Friedmann
equation, has to be added as an additional  constraint.

\section{Validity of the four-dimensional approach ?}

In this section we  analyze the coupled dynamics of the scale factor   and
radion field that one would naively deduce from  the four--dimensional
action~(\ref{action}) considering the three  functions $N(t)$, $a_0(t)$ and
$\R(t)$ as dynamical variables for the  variational problem. This means  that
we  do not impose eq.~(\ref{fried0}) as an  external constraint. 

The corresponding system is analogous to that of the   scale factor and a 
scalar field in four-dimensional FLRW cosmology,  their dynamics being
determined by a coupled system of second  order differential equations, in
addition to the constraint  which comes from the variation of the action with
respect to $N$. Thus, at a fiducial initial time $t_*$, one  must specify, for
example,  the values of $a_0\left(t_*\right)$ (which is in fact arbitrary
because  of the rescaling property of the system), $\R\left(t_*\right)$ and
$\dot \R\left(t_*\right)$ while $\dot a_0\left(t_*\right)$ is deduced from  the
constraint equation.  This must be contrasted with  the full five--dimensional
dynamics, which is described by  eqs.~(\ref{fried0})~and~(\ref{bdl}), where the
only quantity to be specified at an initial time $t_*$ is 
$\R\left(t_*\right)$, since  $\dot\R\left(t_*\right)$ is  determined by
eq.~(\ref{bdl}). We can thus see that the action (\ref{action})  generates
more  solutions  than the true  (five-dimensional) solutions  which  belong to
a subspace characterized by the unconventional Friedmann equation
(\ref{fried0}). 

Keeping this in mind, let us however examine in more  detail the ``theory''
suggested by the four-dimensional action  (\ref{action}). We will not write
here the second order differential  equations, which are rather cumbersome, 
but the  inspection of  the Friedmann equation~(\ref{friedmann}) is  already
instructive in itself. A first remarkable feature of eq.~(\ref{friedmann}) is
that the energy densities  of the two branes enter {\it linearly} in the
Friedmann equation,  the energy density of the second brane being corrected by
both a  warping effect and a Lorentz factor due to the motion of the brane.
This linear behaviour is of course familiar in ordinary cosmology but  might
appear  more surprising in brane cosmology where the brane energy  density
enters {\it quadratically}. As remarked in the previous section, the correct
behaviour is recovered only if the unconventional Friedmann law is imposed by
hand on one of the two branes.

One can give the Friedmann equation~(\ref{friedmann}) a  more familiar aspect
by introducing  the effective static potential for the radion.  It  can be read
directly from the action (\ref{action}) and its expression is given by 
\beq
\kappa^2\,V_{stat}(\R)= -12\,\mu\,\left(\psi_2+\psi_3\right)
+\kappa^2\,\sigma_0+\kappa^2\,\sigma_1\,\A_1^4.
\label{V_stat}
\eeq
The total effective potential for the radion can then be deduced by 
including gravity. It is given by the expression
\beq
\kappa^2\,V_{tot}(\R)= \kappa^2\,V_{stat}- 12\,{H_0^2\over \mu}\,\psi_1.
\eeq
One can check that its extremum yields  the equilibrium position  $\R_{eq}$
defined above in (\ref{equilibrium}).

Using the static potential (\ref{V_stat}), one  can now rewrite the Friedmann
equation (\ref{friedmann}) in the  form 
\beq
3\,H_0^2=\kappa_4^2\left[V_{stat}-\frac{6}{\kappa^2}\left(H_0+\mu\,{\A_1'\over \A_1}\dot \R\right){\A_1^2\,\dot\R\over 1-(\dot\R/\A_1)^2}
-\sigma_1\,\A_1^4\,\left(1-{1\over 
\sqrt{1-(\dot\R/\A_1)^2}}\right)\right], 
\label{friedmann2}
\eeq 
where the effective four--dimensional gravitational coupling is defined by  
\beq
\kappa_4^2 \equiv
\frac{\kappa^2\,\mu}{2\,\psi_1\left(\R\right)}. 
\label{kappa_4}
\eeq 
Let us now consider an expansion in the parameter $\dot\R/\A_1$, which means
that  the velocity of the radion  measured by an observer on the second brane
is small. Keeping terms only up to $\dot\R^2$ on the right hand side,  one then
finds in addition to the potential,  a  kinetic contribution  for the radion,
which reads 
\beq
\kappa_4^2\left[-\frac{12}{\kappa^2}\mu \,\A_1'\,\A_1
+\sigma_1\,\A_1^2\,\right]\frac{\dot \R^2}{2}
\eeq
as well as  a coupling between the radion velocity and the Hubble parameter.
The latter  term can be understood by observing that the  action~(\ref{action})
describes a Brans-Dicke type theory. In such theories, the scalar curvature in
the Lagrangian is multiplied by a scalar field, say $\Psi$. Then an extra term
of the form  $-3\,H_0\, \dot\Psi/\Psi$ appears on the right-hand side of the
Friedmann equation. In our case, $\Psi=\psi_1$ and $\dot
\psi_1=\A_1^2\,\dot\R$, which accounts for the term that appears in
eq.~(\ref{friedmann2}).  With the above expansion, the Friedmann equation
acquires a  familiar  look, but in the more general case where the  radion
velocity is not assumed to be small, all the terms  are modified  by 
corrections due to the Lorentz factor $(1-(\dot\R/\A_1)^2)^{-1/2}$.
Nevertheless, even when the radion velocity is small, the four dimensional
system yields a space of solutions that is larger than the actual space of
solutions of the full five--dimensional dynamics.

In fact, things are slightly more subtle  because the most  general Friedmann
equation resulting from the five-dimensional analysis is not (\ref{fried0}),
but includes  a radiation-like Weyl term as well  (see eq.~(\ref{fried}))
with   $\C$ as an integration constant. One can thus wonder if the extra
freedom among the initial  conditions of the four-dimensional system can 
somehow mimic this  constant of integration of the five-dimensional dynamics.
It turns out that this is the case  in the  particular Randall-Sundrum limit
(i.e. with critical branes,  $\eta_0=1$ and $\eta_1=-1$) and in the 
slow-velocity approximation, as  was shown explicitly in  \cite{kost01} by
deriving an effective action for the  moduli consisting of the scale factors of
the two branes. 

Here, we can reconsider this question from a more general point of view.  To
compare the four-dimensional dynamics with the true five-dimensional dynamics,
we consider the quantity 
\beq
\chi\equiv \dot H_0+2\left[H_0^2-\mu^2\,\left(\eta_0^2-1\right)\right]\,\,.
\eeq
When the brane energy density is constant, as is the case here,  this quantity
is exactly zero for the true five-dimensional dynamics~(\ref{fried}), even 
when one has a radiation-like term (i.e. $\C\neq 0$ in (\ref{fried})).   The 
value of $\chi$ obtained from the four-dimensional system will therefore
represent directly the deviation from  the true dynamics.

Let us first consider the Randall-Sundrum regime,  for which $\eta_0=1$ and
$\eta_1=-1$.  Expanding with respect  to  the  brane velocity $\dot\R$, one
obtains from the equations of motion that 
\beq
\dot H_0+2\,H_0^2={3+4\,\A_1+\A_1^2\over 2\,(1-\A_1)^2}\,\mu^2\dot\R^4 
+ {\cal O}\left(\dot\R^6\right)\,\,.
\eeq
The above formula shows that, at order ${\cal{O}}(\dot\R^2)$, the 
four-dimensional effective action yields the expansion law of a 
radiation--dominated universe: the constant $\C$ is in fact  mimicked by a
constant term  proportional to $\dot{\R}(t_*)^2$.

There is another regime where the four-dimensional dynamics can  mimic the true
five-dimensional dynamics. This is when the radion is near  its equilibrium
point, defined by  (\ref{equilibrium}).  Keeping only terms up to second order
in time derivatives in the equations of motion, one finds
\beq
\chi=\mu^2\, {(\eta_0^2-1)\,\dot\R^2\over \A_1^2-2\,\eta_1\,\psi_1}.
\eeq
The four-dimensional  dynamics thus approximates the true dynamics near  the
equilibrium point if
\beq
\epsilon_0\,\dot \R^2 \ll H_0^2/\mu^2,
\eeq
with $\epsilon_0=\eta_0-1\ll 1$.

The analysis of the equilibrium regime also gives a solution  to the  apparent
contradiction, mentioned in the introduction, between the Brans-Dicke type
behaviour of the effective four-dimensional theory (corroborated by  the
analysis of the fluctuations \cite{gt99}) and  the unconventional Friedmann
equations, which are independent of the  radion, i.e. of the Brans-Dicke field
(see also a recent discussion on this problem in \cite{mk02} from a different
perspective). This can be seen in the low energy limit $\epsilon_0\ll 1$, where
one gets approximately the usual Friedmann equation with 
\beq
\tilde\kappa_4^2=\kappa^2\,\mu,
\eeq 
in contrast with the gravitational coupling found in (\ref{kappa_4}) which
depends explicitely on the radion. 

The explanation of this paradox comes from the observation that, in the
effective theory, Brans-Dicke gravity couples to the energy densities on both
branes,  as is explicit in the Friedmann equation (\ref{friedmann2}), which 
for $\dot\R=0$ reduces to 
\beq
3\,H_0^2=\kappa_4^2\,\,V_{stat},
\eeq
with the potential $V_{stat}$ containing an average sum of the energy 
densities. When the system is close to the Randall-Sundrum regime, one finds that 
\beq
\kappa^2\,V_{stat}=6\,\mu\,\left(\epsilon_0+\epsilon_1\,e^{-4 \mu\R}\right)\,\,,
\eeq
where we have defined $\epsilon_1\equiv \eta_1+1\ll 1$. As seen above, the
four-dimensional effective theory approximates well the true five--dimensional
dynamics only close to the equilibrium point, which requires a fine--tuning
between the tension on the two branes, the relation being
\beq
\epsilon_1=-\epsilon_0\,e^{2\mu\R}.
\eeq
This coupling between $\epsilon_1$ and $\epsilon_0$ exactly cancels the 
dependence of $\kappa_4^2$ on the radion, and the gravitational coupling 
to only the matter in our brane does not depend on the radion as expected.

\section{Conclusions}

We have computed the four-dimensional effective action, which governs the  full
non-perturbative dynamics of the homogeneous radion.  The correct dynamics for
the  radion, including corrections due to the Lorentz factor,  is obtained in
all cases only if one imposes ``by hand'' the  non-conventional Friedmann
equation  for the brane scale factor. 

Our work  emphasizes the fact that one cannot in general use a 
four-dimensional approach  to describe a setup which is  intrinsically
five-dimensional, even if our universe is, in this type of  model, a
four-dimensional manifold where ordinary matter is confined.  This is
reminiscent of the realization that the Friedmann equation in the  brane must
include a $\rho^2$ term because of the five-dimensional nature  of gravity
\cite{bdl99}. Noticing that,  in the case of the Friedmann equation, one
recovers the usual four-dimensional  form at low energy, i.e. when our
brane-universe is close to Minkowski, one could argue that the four-dimensional
and five-dimensional approaches are equivalent at low energies. However, this
can  sometimes be  misleading as shown in this work, since at low energy, in a
two-brane system, one requires to fine--tune the brane tensions in such a way
that gravity looks purely tensorial rather than scalar--tensor if one demands
the four--dimensional effective cosmological dynamics to reproduce the actual
five-dimensional one.

In the present work, by considering a  system simple enough but 
yielding a non trivial effective potential for the radion,  we have been able
to compare explicitly the effective and true dynamics. The  space of effective
solutions is larger than the space of true  solutions corresponding to the
five-dimensional ansatz, because one constraint is lost in the integration of
the action over  the fifth dimension. We have shown that in a  restrictive
range of parameters, namely  for critical branes or for a radion near
equilibrium point, and  with a  slow radion in both cases, the two dynamics are
compatible (in these two regimes, the effective solutions coincide with  the
enlarged space of five-dimensional solutions which allow  for Weyl radiation
even if there solutions were not included in the  ansatz before integration of
the action). We were  also  able to evaluate quantitatively  the deviation of
the effective dynamics  with respect to the true dynamics. 

By extension, our analysis suggests that studying brane cosmology in a
five-dimensional setup appears unavoidable as soon as one wishes to explore
regimes far from the quasi-static limit. And, indeed, these regimes  might
produce new and potentially  interesting effects, which cannot be seen in the
already thoroughly  explored four-dimensional models. 

\acknowledgements

We would like to thank P. Bin\'etruy and C. Deffayet 
for very instructive discussions, N. Deruelle and  G. Veneziano for 
reading and 
commenting our manuscript.  D.L. would like to thank the organizers
of the workshop on ``Braneworld- Dynamics of spacetime with boundary'' 
(YITP, Kyoto), where he had stimulating discussions 
on the present work, in particular with J. Garriga, N. Kaloper, R. Maartens, 
K. Maeda, S. Mukohyama, M. Sasaki, T. Tanaka.


\begin{thebibliography}{99}

\bibitem{rs99a}
L.~Randall and R.~Sundrum,
Phys.\ Rev.\ Lett.\  {\bf 83} (1999) 3370
[arXiv:hep-ph/9905221].

\bibitem{rs99b}
L.~Randall and R.~Sundrum,
Phys.\ Rev.\ Lett.\  {\bf 83} (1999) 4690
[arXiv:hep-th/9906064].

\bibitem{gt99} 
J.~Garriga and T.~Tanaka,
Phys.\ Rev.\ Lett.\  {\bf 84} (2000) 2778
[arXiv:hep-th/9911055].

\bibitem{gw99} 
W.~D.~Goldberger and M.~B.~Wise,
Phys.\ Rev.\ Lett.\  {\bf 83} (1999) 4922
[arXiv:hep-ph/9907447].

\bibitem{bdel99} 
P.~Bin\'etruy, C.~Deffayet, U.~Ellwanger and D.~Langlois,
Phys.\ Lett.\ B {\bf 477} (2000) 285
[arXiv:hep-th/9910219].

\bibitem{cgr99} 
C.~Charmousis, R.~Gregory and V.~A.~Rubakov,
Phys.\ Rev.\ D {\bf 62} (2000) 067505
[arXiv:hep-th/9912160].

\bibitem{radion}
Z.~Chacko and P.~J.~Fox,
Phys.\ Rev.\ D {\bf 64} (2001) 024015
[arXiv:hep-th/0102023]~,
U.~Gen and M.~Sasaki,
arXiv:gr-qc/0201031.

\bibitem{bdl01}
P.~Bin\'etruy, C.~Deffayet and D.~Langlois,
Nucl.\ Phys.\ B {\bf 615} (2001) 219
[arXiv:hep-th/0101234].


\bibitem{kkl01}
R.~Kallosh, L.~Kofman, A.~D.~Linde and A.~A.~Tseytlin,
Phys.\ Rev.\ D {\bf 64} (2001) 123524
[arXiv:hep-th/0106241].

\bibitem{kost01} 
J.~Khoury, B.~A.~Ovrut, P.~J.~Steinhardt and N.~Turok,
Phys.\ Rev.\ D {\bf 64} (2001) 123522
[arXiv:hep-th/0103239].

\bibitem{kraus} 
P.~Kraus,
JHEP {\bf 9912} (1999) 011
[arXiv:hep-th/9910149].

\bibitem{gh}
G.~W.~Gibbons and S.~W.~Hawking,
Phys.\ Rev.\ D {\bf 15} (1977) 2752.


\bibitem{mk02}
S.~Mukohyama and L.~Kofman,
arXiv:hep-th/0112115.


\bibitem{bdl99}
P.~Bin\'etruy, C.~Deffayet and D.~Langlois,
Nucl.\ Phys.\ B {\bf 565} (2000) 269
[arXiv:hep-th/9905012].

\end{thebibliography}
\end{document}